\documentclass[sigconf,screen]{acmart}

\newcommand{\final}{1}

\usepackage{balance} 

\usepackage{enumitem} \usepackage{booktabs}
\usepackage[skip=1pt,labelfont=bf]{caption}
\usepackage{multirow}
\usepackage{framed}
\usepackage[flushleft]{threeparttable}
\usepackage{makecell}
\usepackage{listings}
\usepackage{pifont}
\usepackage{microtype}

\definecolor{codegreen}{rgb}{0,0.6,0}
\definecolor{codegray}{rgb}{0.5,0.5,0.5}
\definecolor{codepurple}{rgb}{0.58,0,0.82}
\definecolor{codered}{rgb}{1,0.6,0.6}
\definecolor{backcolour}{rgb}{0.95,0.95,0.92}
\definecolor{lightgray}{gray}{0.9}

\lstdefinestyle{mystyle}{
commentstyle=\color{codegreen},
    keywordstyle=\color{magenta},
    numberstyle=\tiny\color{codegray},
    stringstyle=\color{codepurple},
    basicstyle=\footnotesize,
    breakatwhitespace=false,
    breaklines=true,
    captionpos=b,
    keepspaces=true,
showspaces=false,
    showstringspaces=false,
    showtabs=false,
    tabsize=2
}

\newboolean{showcomments}
\setboolean{showcomments}{true}
\ifthenelse{\boolean{showcomments}}
 { \newcommand{\mynote}[2]{
      \fbox{\bfseries\sffamily\scriptsize#1}
        {\small$\blacktriangleright$\textsf{\emph{#2}}$\blacktriangleleft$}}}
        { \newcommand{\mynote}[2]{}}

\definecolor{SithColor}{rgb}{0.7,0,0}

\newcommand{\ra}[1]{\renewcommand{\arraystretch}{#1}}

\ifthenelse{\equal{\final}{1}}
{
}
{}

\definecolor{diffstart}{RGB}{173, 175, 177}
\definecolor{diffincl}{RGB}{0, 110, 0}
\definecolor{diffrem}{RGB}{255, 50, 0}
\definecolor{orange}{RGB}{255, 165, 0}
\definecolor{javapurple}{rgb}{0.5,0,0.35} 

\AtBeginDocument{}

\setcopyright{acmlicensed}
\acmPrice{15.00}
\acmDOI{10.1145/3597926.3598080}
\acmYear{2023}
\copyrightyear{2023}
\acmSubmissionID{issta23main-p151-p}
\acmISBN{979-8-4007-0221-1/23/07}
\acmConference[ISSTA '23]{Proceedings of the 32nd ACM SIGSOFT International Symposium on Software Testing and Analysis}{July 17--21, 2023}{Seattle, WA, United States}
\acmBooktitle{Proceedings of the 32nd ACM SIGSOFT International Symposium on Software Testing and Analysis (ISSTA '23), July 17--21, 2023, Seattle, WA, United States}
\received{2023-02-16}
\received[accepted]{2023-05-03}

\begin{document}

\title{Towards More Realistic Evaluation for Neural Test Oracle Generation}

\author{Zhongxin Liu}
\affiliation{
  \institution{Zhejiang University}
  \country{China}
}
\email{liu_zx@zju.edu.cn}

\author{Kui Liu}
\authornote{Corresponding author.}
\affiliation{\institution{Huawei}
\country{China}
}
\email{brucekuiliu@gmail.com}

\author{Xin Xia}
\affiliation{
  \institution{Huawei}
  \country{China}
}
\email{xin.xia@acm.org}

\author{Xiaohu Yang}
\affiliation{
  \institution{Zhejiang University}
  \country{China}
}
\email{yangxh@zju.edu.cn}

\begin{abstract}
Unit testing has become an essential practice during software development and maintenance. Effective unit tests can help guard and improve software quality but require a substantial amount of time and effort to write and maintain. A unit test consists of a test prefix and a test oracle. Synthesizing test oracles, especially functional oracles, is a well-known challenging problem. Recent studies proposed to leverage neural models to generate test oracles, i.e., neural test oracle generation (NTOG), and obtained promising results.

However, after a systematic inspection, we find there are some inappropriate settings in existing evaluation methods for NTOG. These settings could mislead the understanding of existing NTOG approaches' performance. We summarize them as \ding{172} generating test prefixes from bug-fixed program versions, \ding{173} evaluating with an unrealistic metric, and \ding{174} lacking a straightforward baseline. In this paper, we first investigate the impacts of these settings on evaluating and understanding the performance of NTOG approaches. We find that \ding{182} unrealistically generating test prefixes from bug-fixed program versions inflates the number of bugs found by the state-of-the-art NTOG approach TOGA by 61.8\%, \ding{183} FPR (False Positive Rate) is not a realistic evaluation metric and the Precision of TOGA is only 0.38\%, and \ding{184} a straightforward baseline NoException, which simply expects no exception should be raised, can find 61\% of the bugs found by TOGA with twice the Precision. Furthermore, we introduce an additional ranking step to existing evaluation methods and propose an evaluation metric named Found@K to better measure the cost-effectiveness of NTOG approaches in terms of bug-finding. We propose a novel unsupervised ranking method to instantiate this ranking step, significantly improving the cost-effectiveness of TOGA. Eventually, based on our experimental results and observations, we propose a more realistic evaluation method TEval+ for NTOG and summarize seven rules of thumb to boost NTOG approaches into their practical usages.
\end{abstract}



\maketitle

\section{Introduction}\label{sec:intro}

Unit testing has become an accepted and even mandatory practice during software development.
For a component (a method, class, or module), its unit tests check whether its implemented functionality can match its intended functionality and document its intended usage.
Developers leverage unit tests to find bugs, identify regressions, and facilitate the understanding and usage of the corresponding components.
Therefore, effective unit tests can guard and improve software quality as well as reduce the costs of software failures~\cite{hartman2002issta, planning2002economic}.

However, writing high-quality unit tests is non-trivial and time-consuming.
Prior work has shown that developers spend more than 15\% of their time in writing tests~\cite{daka2014survey}.
To address this challenge, some tools are proposed to automate unit test generation~\cite{pacheco2007feedbackdirecteda,fraser2012whole,lukasczyk2020automated}.
A unit test consists of a test prefix, which is a sequence of statements driving the unit under test into a specific state~\cite{dinella2022toga}, and a test oracle, which specifies the condition that should be satisfied in such state~\cite{dinella2022toga}.
For example, in the unit test presented in Figure~\ref{fig:test_case}, lines 3-6 are the test prefix, which creates a KeyedValues object, inserts an item into the object, and then removes the item.
Line 7 specifies the oracle, i.e., the object should contain no item after executing the test prefix.
Automated unit test generation tools aim at generating both test prefixes and oracles~\cite{pacheco2007feedbackdirecteda,fraser2012whole}.
However, since these tools are unknown of the intended behavior of a generated test prefix, they either generate regression oracles~\cite{xie2006augmentinga}, i.e., regarding the observed behavior as the oracle, or use implicit oracles~\cite{barr2014oracle}, such as ``program crashes are usually undesirable''.
Regression oracles cannot reveal functional bugs in the project versions where they are generated.
Implicit oracles are not always correct and hence are neither sufficient nor accurate to find the violations of intended functionality.
To sum up, these automated unit test generation tools still can not replace manual test writing.

\begin{figure}[!t]
    \centering
    \vspace{3mm}
\begin{lstlisting}[language=Java,linewidth={\linewidth},frame=tb,basicstyle=\small\ttfamily]
@Test
public void test14()  throws Throwable  {
   KeyedValues KeyedValues0 = new KeyedValues();
   Integer integer0 = new Integer(233);
   KeyedValues0.setValue((Comparable) integer0, 0.0);
   KeyedValues0.removeValue((Comparable) integer0);
   assertEquals(0, KeyedValues0.getItemCount());
}
\end{lstlisting}
    \caption{Example of a unit test case.}
    \vspace{-0.3cm}
    \label{fig:test_case}
\end{figure}

To tackle the oracle problem~\cite{barr2014oracle} and complement unit test generation tools, researchers proposed several approaches to automatically generate test oracles and achieved promising results~\cite{pandita2012inferring,tan2012tcomment,goffi2016automatic,blasi2018translating,zhai2020c2s,watson2020learning,tufano2022generating, dinella2022toga}.
Recently, a popular and effective category of approaches trains neural networks on developer-written unit tests to learn to generate test oracles~\cite{watson2020learning,tufano2022generating, dinella2022toga}, namely neural test oracle generation (\textbf{NTOG}) approaches.
Given a focal unit and a test prefix, NTOG approaches leverage the trained neural networks to either generate an oracle from scratch or select an oracle from several candidates that are pre-generated based on templates.
NTOG approaches are flexible since they do not rely on natural language patterns and manual-crafted rules.
The state-of-the-art NTOG approach is TOGA proposed by Dinella et al.~\cite{dinella2022toga}, which is shown to outperform existing test oracle generation tools in terms of bug-finding.
Given a test prefix and its focal context (including the focal method and the focal docstring), TOGA first generates a set of possible oracles based on several empirically summarized templates and the type-based constraints inferred from the test prefix, and then fine-tunes two CodeBERT models to rank the generated candidates and outputs the oracle ranking first.

A fundamental goal of test oracle generation is to uncover bugs.
To evaluate the bug-finding performance of NTOG approaches, two evaluation methods are widely adopted by prior work.
The first one is to evaluate the target approach in a held-out set of developer-written unit tests collected from real-world projects~\cite{watson2020learning,mastropaolo2021studying,dinella2022toga}.
This is a common evaluation method for neural methods and is also used in other code-related tasks, such as code comment generation~\cite{hu2018deep, hu2019deep} and neural patch generation~\cite{chen2019sequencer,tufano2019learning}.
This method can reflect whether one NTOG approach is better than other ones in terms of learning ability.
However, it assumes the existence of developer-written test prefixes, which is usually not the case in practice, and thus cannot explicitly demonstrate the effectiveness of NTOG approaches in uncovering real-world bugs.
Another evaluation method is proposed in the TOGA paper~\cite{dinella2022toga} and we refer to it as \textbf{TEval} for convenience.
TEval first leverages a unit test generation tool, such as EvoSuite~\cite{fraser2012whole}, to generate test prefixes for NTOG approaches, then combines the oracles generated by NTOG approaches with the corresponding test prefixes to construct complete test cases, and finally checks whether the generated oracles can find bugs by compiling and executing such test cases.
Because TEval does not require developer-written test prefixes, it is more realistic than the first evaluation method and can measure the performance of NTOG approaches in uncovering real-world bugs.

Using TEval, Dinella et al.~\cite{dinella2022toga} presented that TOGA can find 57 out of 835 bugs in the Defects4J benchmark~\cite{just2014defects4j} with a False Positive Rate (FPR) of 25\%.
After a systematic inspection, we find that even TEval is not realistic enough and could not accurately assess the real-world performance of NTOG approaches.
Because some of its settings are inappropriate and 
could mislead the understanding of existing NTOG approaches' performance.
We summarize them into three aspects below:

\textbf{\ding{182} Generating test prefixes from bug-fixed program versions.}
Existing NTOG approaches take as input test prefixes and focal context (e.g., focal methods and their docstrings).
To generate test prefixes for a bug,
TEval runs EvoSuite on the corresponding bug-fixed program version.
However, when applying NTOG tools in practice, we expect them to find bugs in the current buggy program version, and the bug-fixed program version is not available at that time.
Therefore, to be consistent with the real-world usage scenario, test prefixes should be generated on the buggy program version instead of the bug-fixed program version.
Also, considering that EvoSuite is guided by branch coverage, the test prefixes generated by EvoSuite on the bug-fixed program version may implicitly contain some information related to the patch of the bug, which should be unknown during bug finding.
Therefore, using such test prefixes as input may result in information leakage and inflate the performance metrics.

\textbf{\ding{183} Evaluating with an unrealistic metric.}
TEval uses FPR (False Positive Rate) to assess and compare against different NTOG approaches.
FPR is calculated by $\#FP / (\#FP + \#TN)$, where FP (False Positive) refers to the generated oracle that fails on both the buggy and the bug-fixed program versions, TN (True Negative) refers to the generated oracle that passes on both versions, and \#FP and \#TN denote the number of FPs and TNs, respectively.
However, when using bug-finding tools, developers care more about the Precision, i.e., how many FPs they need to inspect to find a bug, and usually have few interests in using a tool with a low Precision~\cite{bessey2010fewa, johnson2013why}.
Therefore, we argue that the FPR metric is not consistent with developers' concerns and may not accurately reflect the real-world performance of NTOG approaches.

\textbf{\ding{184} Lacking a straightforward baseline.}
According to TOGA's evaluation results~\cite{dinella2022toga}, the observed behaviors of 39 out of the 57 bugs (68.4\%) found by TOGA are ``Exception Raised''.
The oracles of these bugs, i.e., no exception raised, are straightforward and do not need to be explicitly specified by generating an assert statement.
Considering this, we construct a straightforward oracle generation approach named NoException, which expects that every test prefix does not raise exceptions and does not explicitly generate assert statements.
Comparing NTOG approaches with NoException on bug benchmarks can help practitioners understand how much they can gain from neural models compared to implicit oracles.
However, prior work has not explored this straightforward baseline yet.

To boost neural test oracle generation (NTOG) towards more realistic evaluation, in this paper, we first investigate the effects of the inappropriate settings in TEval on evaluating and understanding the performance of NTOG approaches.
Specifically, we conduct a case study using TOGA~\cite{dinella2022toga}, the latest and state-of-the-art NTOG approach.
We address the three inappropriate settings in TEval one by one and re-evaluate TOGA on the Defects4j benchmark using the improved TEval.
We observe that: 
\ding{172} Generating test prefixes from the bug-fixed program versions significantly inflates the performance metrics. On average, it improves the number of bugs found by TOGA from 64.4 to 104.2 and enhances FPR from 22.1\% to 19.7\% (the lower, the better). 
\ding{173} The Precision of TOGA is only 0.38\%, which means developers need to check over 260 test cases that fail on the buggy program version to find a bug-revealing one.
\ding{174} On average, 61\% of the bugs found by TOGA can be found by the straightforward baseline NoException with twice the Precision, and on average 4.7 bugs found by NoException cannot be found by TOGA.

Considering the limited development resources, we argue that when applying an NTOG approach, it is impractical to ask developers to check all the generated test cases that fail if most of them cannot reveal bugs.
The bug-finding performance of NTOG approaches should be measured in a cost-effective way.
Therefore, we further enhance TEval by introducing an additional ranking step and evaluating NTOG approaches' performance by counting the number of bugs that can be found if developers only check the top-k ranked test cases for each bug, i.e., Found@K.
In addition, we propose a novel unsupervised ranking method, which leverages a set of features designed for this task and an outlier detection algorithm, to instantiate this ranking step.
Experimental results show that \ding{172} our ranking method can significantly improve random ranking in terms of Found@K, and \ding{173} TOGA can be more cost-effective than NoException with proper ranking.

Finally, we propose a more realistic evaluation method named TEval+ for NTOG by incorporating all our improvements with TEval, and further summarize seven rules of thumb to boost NTOG approaches towards their practical usage.

Our main contributions are summarized as follows:

\begin{itemize}
    \item We point out several inappropriate settings in existing evaluation methods for NTOG and investigate their effects on measuring and understanding the bug-finding performance of the state-of-the-art NTOG approach TOGA.
    \item We propose an evaluation method named TEval+ for NTOG, which enhances TOGA's evaluation method by addressing its inappropriate settings, reducing duplicates and noise during evaluation, and introducing an additional ranking step to rank failed test cases. We believe TEval+ is more realistic than existing evaluation methods and can better reflect the bug-finding performance of NTOG approaches.
    \item We proposed a novel unsupervised method to instantiate the ranking step of TEval+. Experimental results show that it can significantly improve the cost-effectiveness of NTOG approaches.
    \item We summarize seven rules of thumb to help practitioners conduct more realistic evaluations for NTOG approaches.
    \item Our replication package can be found at~\cite{website:package}.
\end{itemize}

The remainder of this paper is organized as follows: 
Section~\ref{sec:background} introduces the Defects4J benchmark and existing evaluation methods for NTOG.
Section~\ref{sec:approach} investigates the effects of inappropriate evaluation settings on evaluating and understanding the performance of NTOG approaches.
We elaborate on our unsupervised ranking algorithm and its evaluation results in Section~\ref{sec:ranking}.
Section~\ref{sec:pipeline} describes our enhanced evaluation method TEval+ and our summarized rules of thumb for NTOG.
In Section~\ref{sec:discussion}, we present TOGA's performance on different types of oracles and discuss the threats to validity.
After reviewing the related work in Section~\ref{sec:related_work}, we conclude and point out future work in Section~\ref{sec:conclusion}.
 \section{Background}\label{sec:background}
This section briefly introduces the Defects4J benchmark~\cite{just2014defects4j} and existing evaluation methods for NTOG.

\subsection{The Defects4J Benchmark}\label{sec:defects4j}
Defects4J~\cite{just2014defects4j} is a widely-used bug benchmark containing 835 bugs from 17 real-world Java projects.
TEval uses Defects4J to evaluate the bug-finding performance of NTOG approaches.
Each bug collected by Defects4J is recorded in the corresponding issue tracker, is fixed in a single commit by modifying the source code, and includes a buggy and a bug-fixed program versions.
Each bug-fixed program version is based on a manually minimized patch and passes all test cases, while each buggy program version fails at least one test case and such test cases are deterministic.
Defects4J also provides a supporting framework to evaluate a test suite on a specific buggy/bug-fixed program version.
Because the only difference between a buggy program version and its corresponding bug-fixed program version is a minimal patch fixing the bug, a test case that fails on the buggy version and passes on the bug-fixed version must trigger the specific bug.
This property can be used to determine whether a generated test case is a bug-finding one.
Following Dinella et al.~\cite{dinella2022toga}, we also use Defects4J 2.0.0 in this work.

\subsection{Evaluation Methods for Neural Test Oracle Generation}\label{sec:eval_methods}
Existing NTOG approaches take as input a test prefix and the corresponding focal context (e.g., the focal method and/or its docstring comment) and output an assert statement, i.e., an oracle.
A common way to evaluate NTOG approaches is to execute them on a held-out test set collected from developer-written unit tests.
In detail, for each test sample, such evaluation method extracts its test prefix and focal context as input, uses the target NTOG approach to generate an oracle, and regards the corresponding developer-written oracle as ground truth to measure the generation accuracy.
However, when applying NTOG approaches in practice, developer-written test prefixes are usually not available, and writing high-quality or bug-reaching test prefixes is usually not easier or even more difficult and time-consuming than writing test oracles for developers.
Therefore, it is impractical and somehow unrealistic to assume the existence of developer-written test prefixes when evaluating NTOG approaches.
Thus, such evaluation method can hardly reflect the real-world bug-finding performance of NTOG approaches.

Tufano et al.~\cite{tufano2022generating} proposed to evaluate NTOG approaches by generating additional assert statements for the test cases generated by EvoSuite, inserting such asserts as the last statements of the test cases, and calculating the differences in code coverage before and after the insertion.
However, their evaluation only chose one bug from Defects4J, required to manually select the focal methods, the generated test cases and the generated asserts, and did not measure the bug-finding performance of NTOG approaches.

Recently, an evaluation method is proposed by Dinella et al.~\cite{dinella2022toga} to assess the real-world bug-finding performance of NTOG approaches.
We refer to it as TEval (\textbf{T}OGA's \textbf{Eval}uation method).
TEval does not assume the existence of developer-written test prefixes.
Instead, it generates test prefixes using an automated unit test generation tool named EvoSuite~\cite{fraser2012whole}.
Therefore, it is more realistic than the first evaluation method.
Specifically, TEval first uses EvoSuite with branch coverage as the criterion on the bug-fixed program versions to generate regression tests.
Then, it extracts the test prefix and the corresponding focal context of each generated test case and feeds them into an NTOG approach for oracle generation.
Each generated oracle is appended to its corresponding test prefix to construct a complete test case.
Finally, it executes each generated test case on the corresponding buggy and bug-fixed program versions.
If a test case fails on the buggy version but passes on the bug-fixed one, it is regarded as a bug-finding test.

Following the definitions in the TOGA paper~\cite{dinella2022toga} and based on the execution results of the generated tests, the generated tests can be divided into four groups: True Positive (TP), True Negative (TN), False Positive (FP) and False Negative (FN).
Positive and Negative mean that the generated test case fails and passes on the corresponding buggy program version, respectively.
True/False denotes that the generated test passes/fails on the corresponding bug-fixed program version.
Based on these definitions, a bug-finding test is a TP sample.
To measure the bug-finding performance, TEval uses two metrics, i.e., BugFound and FPR (False Positive Rate).
BugFound refers to the number of bugs that can be uncovered by TP samples.
FPR is calculated by $\#FP/(\#FP + \#TN)$.
The experimental results of using TEval on TOGA show that TOGA, the state-of-the-art NTOG approach, can find 57 bugs on Defects4J with an FPR of 25\%, which is impressive and outperforms all baseline approaches, including seq2seq~\cite{tufano2022generating} and JDoctor~\cite{blasi2018translating}, by substantial margins.
 \section{The Impact of Inappropriate Settings in TEval}\label{sec:approach}
TEval is currently the state-of-the-art method for evaluating the bug-finding performance of NTOG approaches.
However, after a systematic inspection, we find that TEval also has several inappropriate settings, which may introduce significant gaps between the reported evaluation results and the real-world bug-finding performance of NTOG approaches.
As described in Section~\ref{sec:intro}, we summarize these inappropriate settings as: \textbf{\ding{182} generating test prefixes from bug-fixed program versions}, \textbf{\ding{183} evaluating with an unrealistic metric,} and \textbf{\ding{184} lacking a straightforward baseline.}
In this section, we briefly review each inappropriate setting and investigate its impact on evaluating and understanding the bug-finding performance of an NTOG approach.
We use the state-of-the-art NTOG approach TOGA as the evaluation subject, and aim to answer the following research questions (RQs):

\begin{itemize}[leftmargin=*]
    \item \textbf{RQ1: What is the impact of generating test prefixes from bug-fixed program version on TOGA's performance?}
    \item \textbf{RQ2: How effective is TOGA when we use more realistic evaluation metrics?}
    \item \textbf{RQ3: How effective is TOGA when compared with a straightforward baseline?}
\end{itemize}

\subsection{Experimental Setup}
Following TEval~\cite{dinella2022toga}, we use EvoSuite with branch coverage as the criterion to generate test prefixes for TOGA.
Specifically, we leverage the \texttt{gen\_tests.pl} script provided by Defects4J to automate such generation.
We run EvoSuite for 3 minutes per tested program.

In our experiments, we do not need or construct explicit ground truth.
Because TP, FP, TN and FN are \textbf{not} identified by comparing the generated test oracles with explicit ground truth.
Instead, for each generated test oracle, we combine it with its test prefix to construct a complete test case and execute this test case on both the buggy and the bug-fixed program versions.
TP, FP, TN and FN are then identified strictly based on their definitions (described in Section~\ref{sec:eval_methods}) and the execution results.
For example, if a test case fails on the buggy program version and passes on the bug-fixed program version, its oracle will be regarded as TP.

In addition, the original TEval used in the TOGA paper~\cite{dinella2022toga} only runs EvoSuite once for each bug, which may introduce bias since EvoSuite is based on randomized algorithms~\cite{fraser2012whole}.
To reduce the potential bias, in this work, we run EvoSuite for each bug 10 times with different random seeds and report the mean value of each evaluation metric as the result.

\subsection{RQ1: Generating Test Prefixes from Buggy Versions}
According to the TOGA paper, TEval generates test prefixes ``by running EvoSuite with default settings (i.e., coverage-guided) on the fixed program versions''~\cite{dinella2022toga}.
As discussed in Section~\ref{sec:intro}, this is unrealistic and may result in information leakage and inflate the performance metrics.
In this RQ, we aim to investigate the impact of this unrealistic setting on evaluating TOGA's bug-finding performance and figure out whether it is necessary to fix this setting.

In addition, we find two implementation problems in the official implementation of TEval~\cite{website:toga}.
The first problem is called the \textbf{over-filtering} problem by us.
Specifically, to speed up the execution of test cases, the official implementation aggregates all the test cases generated for a focal class into one test class.
When a test case generated from a buggy/bug-fixed program version is executed on the corresponding bug-fixed/buggy version, it may encounter compilation errors, e.g., using undeclared methods.
If one test case has compilation errors, TEval's official implementation will filter out the whole test class, which could reduce the number of bugs that can be found by an oracle generation approach and underestimate its bug-finding performance.

The other problem is referred to as the \textbf{duplication} problem by us.
In detail, if a test case generated by EvoSuite contains multiple assertions at its end, TEval's official implementation will extract the same test prefix multiple times.
Consequently, the target oracle generation approach will generate duplicate oracles and TEval will construct duplicate test cases.
However, TEval's official implementation does not deduplicate the constructed test cases, which could increase the numbers of TP, FP, TN and FN to varying degrees and therefore bias the evaluation results.

{\bf Approach}: We address the over-filtering problem by identifying and collecting all test cases with compilation errors, removing them from their test classes, and re-executing these test classes to collect execution results.
To solve the duplication problem, for each bug, we deduplicate the constructed test cases before calculating evaluation metrics.
Hereon, we use \textbf{TEval} to refer to our replication of TEval that addresses the duplication problem and keeps other settings the same as the official implementation of TEval.
We refer to the variant of TEval that addresses the two implementation problems and generates test prefixes from bug-fixed program versions as {\bf TEval@fixed}, and the variant which addresses the two problems but generates test prefixes from buggy program versions as {\bf TEval@buggy}.
All three methods remove duplicate test cases since we regard deduplication as a must-do.
Please note that TEval@fixed and TEval@buggy all run EvoSuite 10 times for each bug to generate test prefixes and TEval uses the same test prefixes as TEval@fixed.

To answer this research question, we evaluate TOGA with TEval, TEval@fixed and TEval@buggy.
We illustrate the impact of generating test prefixes from bug-fixed program versions by comparing TEval@fixed with TEval@buggy, and highlight the impact of the over-filtering problem by comparing TEval and TEval@fixed.
In addition, Wilcoxon signed-rank test~\cite{wilcoxon1992individual} at the confidence level of 95\% is used to show whether the more realistic setting significantly affects the evaluation results.
Cliff's delta~\cite{cliff2014ordinal} is used to measure the effect sizes of the performance differences.

\begin{table}[!t]
    \centering
\caption{The performance of TOGA with different evaluation methods.} 
\label{tab:rq1_result}
    \begin{threeparttable}
    \begin{tabular}{@{}lcc@{}}
    \toprule
    \textbf{Evaluation Method} & \textbf{BugFound} & \textbf{FPR}\\
    \midrule
    \textbf{TEval*} & 57 & 25\%\\
    \midrule
    \textbf{TEval} & 96.4 & 17.0\%\\
    \textbf{TEval@fixed} & 104.2 & 19.7\%\\
    \textbf{TEval@buggy} & 64.4 & 22.1\%\\
    \bottomrule
    \end{tabular}
    \begin{tablenotes}
        \footnotesize
        \item The row of TEval* presents the evaluation results reported in the TOGA paper~\cite{dinella2022toga}.
    \end{tablenotes}
    \end{threeparttable}
    \vspace{-0.3cm}
\end{table}

\textbf{Results}:
Table~\ref{tab:rq1_result} presents our experimental results.
The row of TEval* refers to the evaluation results reported by the TOGA paper~\cite{dinella2022toga}.
We can see that our replication of TEval finds 96.4 bugs and achieves an FPR of 17.0\% on average.
Such results are much better than those of TEval*.
Even the experiment with the worst performance can find 86 bugs.
Considering that duplicate test cases cannot help find more bugs, we infer the reason behind the performance gap between TEval and TEval* could be that our hardware environment is quite different from that of TEval* and we generate test prefixes of higher quality.
Comparing TEval with TEval@fixed, we can see that on average TEval@fixed finds more bugs than TEval, which means TEval mistakenly deletes some bug-finding test cases due to the over-filtering problem.
Our statistical analysis shows that the performance differences in terms of BugFound and FPR between TEval and TEval@fixed are significant.
These results indicate that it is necessary to fix the over-filtering problem.

According to Table~\ref{tab:rq1_result}, the evaluation results using TEval@fixed are much better than those using TEval@buggy.
TEval@fixed can find about 61.8\% (39.8) more bugs and reduce the FPR by 2.4\%, although the underlying NTOG approach is the same one.
In addition, our statistical analysis shows that the p-value in terms of BugFound is less than 0.005 and the corresponding effect size is large (i.e., >0.474).
As for FPR, the p-value is less than 0.05 and the effect size is also large.
These results indicate that the differences between the evaluation results using TEval@buggy and TEval@fixed are significant.
Based on such significant differences and considering that only buggy program versions are available in practice, we suggest practitioners use TEval@buggy instead of TEval@fixed for evaluating NTOG approaches.

\vspace{-0.05cm}
\begin{framed}
\vspace{-0.2cm}
In summary, we find that \ding{172} fixing the over-filtering problem can help obtain more accurate evaluation results, and \ding{173} generating test prefixes from bug-fixed program versions can significantly inflate the performance metrics. Considering the practical scenarios of NTOG tools, we believe TEval@buggy can better reflect the real-world performance of NTOG approaches than Teval@fixed.
\vspace{-0.2cm}
\end{framed}

\subsection{RQ2: Evaluating with More Realistic Metrics}\label{sec:approach:rq2}
TEval uses two metrics, i.e., BugFound and FPR, to measure the effectiveness of NTOG approaches in terms of bug finding.
BugFound refers to the total number of bugs that can be revealed by the generated test oracles, which is good for measuring the upper bound (or recall) of an oracle generation approach in terms of bug finding.
FPR is calculated by $\#FP / (\#FP + \#TN)$, as described in Section~\ref{sec:eval_methods}.
Dinella et al. claimed that a high FPR rate implies a developer needs to validate many tests of no use and FPR is a good metric for a bug-finding tool~\cite{dinella2022toga}.
However, when using oracle generation tools in practice, developers only manually check the generated test cases failing on the program, i.e., the positive sample.
TN samples are usually ignored because they pass on the buggy program version and cannot help uncover the hidden bugs.
According to this and the definition of FPR, we argue that {\it FPR cannot reflect the rate of useless test cases that developers need to manually validate, and is not a realistic metric for measuring the bug-finding effectiveness of oracle generation tools}.
In this RQ, we aim to better understand TOGA's bug-finding performance with a more realistic evaluation metric.

\textbf{Approach}: Intuitively, when using oracle generation tools or other bug-finding tools in practice, developers really care about how many FP samples they need to check before finding a TP sample.
In other words, developers value Precision more~\cite{bessey2010fewa, johnson2013why}.
Therefore, we propose to measure the effectiveness and practicability of an NTOG approach by calculating its \textbf{Precision} instead of FPR.
Precision is denoted as $\#TP / (\#FP + \#TP)$, i.e., the ratio of the generated test cases that can indeed find bugs to all the generated test cases failing on the program.
The higher a tool's precision is, the less effort it takes for a developer to validate the generated test cases that is useless.
Like RQ1, we also conduct Wilcoxon signed-rank test~\cite{wilcoxon1992individual} and use Cliff's delta~\cite{cliff2014ordinal} to show whether TEval@fixed significantly outperforms TEval@buggy in terms of Precision.

\begin{table}[!t]
    \centering
\caption{The Performance of TOGA in terms of Precision.} 
\label{tab:rq2_result}
    \begin{threeparttable}
    \begin{tabular}{@{}lcccc@{}}
    \toprule
    \textbf{Evaluation Method} & \textbf{BugFound} & \textbf{Precision} & \textbf{\#TP} & \textbf{\#FP} \\
    \midrule
    \textbf{TEval@fixed} & 104.2 & 1.13\% & 288.0 & 25237.6 \\
    \textbf{TEval@buggy} & 64.4 & 0.38\% & 110.7 & 29099.4\\
    \bottomrule
    \end{tabular}
\end{threeparttable}
    \vspace{-0.2cm}
\end{table}

\begin{figure}[!t]
    \centering
    \vspace{3mm}
\begin{lstlisting}[language=Java,linewidth={\linewidth},frame=tb,basicstyle=\small\ttfamily,numbers=none]
@Test
public void test1() throws Throwable {
   Fraction fraction0 = new Fraction((double) 105, 105);
}

@Test
public void test2() throws Throwable {
   Fraction fraction0 = new Fraction((double) (-290), 542);
}
\end{lstlisting}
    \caption{Two test cases that can both trigger the bug Math1.}
    \vspace{-0.3cm}
    \label{fig:tp_examples}
\end{figure}

\textbf{Result}:
Table~\ref{tab:rq2_result} presents the experimental results.
It is worth mentioning for each evaluation method, the BugFound is smaller than the \#TP, which is expected.
Because for each bug, EvoSuite generates many different test prefixes.
Different test cases can be generated based on such test prefixes, and there can be more than one test case triggering this bug.
For example, the bug Math1 in Defects4J can be triggered by the two different test cases shown in Figure~\ref{fig:tp_examples}.
If an approach generates the two test cases for Math1, its \#TP and BugFound will be increased by 2 and 1, respectively.
Therefore, \#TP can be larger than BugFound.

Surprisingly, the Precision of TOGA using TEval@buggy is only 0.38\% on average, which means that to find a bug-finding test case, developers need to manually check over 260 failed test cases generated by EvoSuite and TOGA.
Even if we unrealistically generate test prefixes from bug-fixed program versions, i.e., using TEval@fixed, TOGA's Precision is still quite low (1.13\%).
Considering the importance of high precision for bug-finding tools, we think there is still a long way to go before directly applying existing NTOG tools in practice.
We also calculate the p-value and the effect size of TEval@buggy compared to TEval@fixed in terms of Precision, which turn out to be less than 0.001 and large.
These results indicate that the Precision of TOGA using TEval@buggy is also significantly worse than that using TEval@fixed, consistent with our findings in RQ1.

\vspace{-0.05cm}
\begin{framed}
\vspace{-0.2cm}
In summary, we find that when using TEval@buggy, TOGA's Precision is only 0.38\%. This calls for more efforts on improving existing NTOG approaches before applying them in practice.
\vspace{-0.2cm}
\end{framed}

\subsection{RQ3: Comparing with a Straightforward Baseline}\label{sec:approach:rq3}
When investigating RQ1 and RQ2, we also manually inspect the generated test oracles to better understand TOGA's effectiveness and practicability.
We find that for the generated test oracles which can find bugs, over 60\% of them simply expect no exception raised.
A similar phenomenon was also observed by Dinella et al.~\cite{dinella2022toga}.
This inspires us that a straightforward baseline, which simply predicts all oracles as no exception raised, may achieve comparable performance.
We refer to this baseline as NoException.
Specifically, NoException does not explicitly generate any assert statement, it directly regards the test prefixes generated by EvoSuite as the generated test cases.
If such a test case throws any exception during execution, NoException regards it as a failed test case.
Comparing NoException with NTOG approaches can help practitioners better understand the benefits they can gain from neural models compared to implicit oracles.
However, to the best of our knowledge, no prior work has explicitly compared existing NTOG approaches with this baseline.
Therefore, this RQ aims to fill this gap.

\textbf{Approach}: 
We use TEval@buggy, which is more realistic than the original TEval, to evaluate NoException and TOGA, and compare their performance in terms of BugFound and Precision.
Also, Wilcoxon signed-rank test~\cite{wilcoxon1992individual} and Cliff's delta~\cite{cliff2014ordinal} are used for statistical analysis.

\begin{table}[!t]
    \centering
    \ra{1.1}
    \caption{Comparison of NoException and TOGA in BugFound and Precision using TEval@buggy.} 
\label{tab:rq3_result}
    \begin{threeparttable}
    \begin{tabular}{@{}lcccc@{}}
    \toprule
    \textbf{Approach} & \textbf{BugFound} & \textbf{Precision} & \textbf{\#TP} & \textbf{\#FP} \\
    \midrule
    \textbf{NoException} &  44.4 & \textbf{0.77\%} & 73.5 & 9415.9 \\
    \textbf{TOGA} & \textbf{64.4} & 0.38\% & 110.7 & 29099.4\\
    \bottomrule
    \end{tabular}
    \end{threeparttable}
    \vspace{-0.3cm}
\end{table}

\textbf{Result}:
Table~\ref{tab:rq3_result} presents the evaluation results of NoException and TOGA.
We observe that NoException can find 44.4 bugs with a Precision of 0.77\% on average.
Although NoException finds 31.1\% (20) fewer bugs than TOGA, its Precision is twice TOGA's Precision.
The statistical analysis also show that NoException's better performance in terms of Precision is significant (p-value < 0.001) with a large effect size.
In addition, we calculate the intersection of the bugs found by NoException and TOGA, and find it contains 39.7 bugs on average.
This means NoException can also find several bugs that cannot be found by TOGA.
These results may imply that the practicability and effectiveness of TOGA require further investigation and discussion.

\vspace{-0.05cm}
\begin{framed}
\vspace{-0.2cm}
In summary, we find that simply generating the oracle that no exception should be thrown can achieve 68.9\% of TOGA's performance in terms of BugFound and twice TOGA's Precision.
\vspace{-0.2cm}
\end{framed}

 \section{An Additional Ranking Step}\label{sec:ranking}
As presented in Section~\ref{sec:approach:rq2}, even for the state-of-the-art NTOG approach TOGA, to find a bug-finding test case, developers need to manually check over 260 failed test cases on average.
This is usually impractical and even unacceptable for developers due to limited development resources in practice.
A more realistic usage scenario is that after an NTOG tool generates many failed test cases for a component, developers only have time to manually check a few of them.
Therefore, an evaluation method for NTOG should also assess the bug-finding performance in a cost-effective way to help practitioners better understand NTOG approaches' practicability.

Motivated by this observation, we first introduce an additional ranking step in TEval@buggy after all generated test cases are executed.
For each bug, before presenting all of its failed test cases to developers, this step ranks these test cases according to their probability of triggering bugs.
Developers can check the top-ranked test cases based on the amount of their available time.
Then, we introduce a new evaluation metric named Found@K, which counts how many bugs can be found if developers only check the top-k recommended test cases for each bug.
It is formally defined as follows:
$$
Found@K = \frac{\sum_{i=1}^{n}\sigma(r_{i} \le K)}{n}
$$
where $r_{i}$ refers to the rank of the first bug-finding test case for bug $i$, $\sigma$ denotes the indicator function, which returns 1 if $r_i \le K$ or 0 otherwise, and $n$ is the number of bugs.
If no bug-finding test case is generated for bug $i$, $r_i$ is set to infinite.
This metric can help measure the cost-effectiveness of NTOG approaches in terms of bug finding, 

\subsection{An Unsupervised Ranking Method}
To instantiate this ranking step, we propose a novel unsupervised ranking method.
Our ranking method is inspired by the following observations:
First, according to the evaluation results in Section~\ref{sec:approach:rq2}, bug-finding test cases (i.e., TPs) are usually rare in all the failed test cases.
Second, bug-finding test cases usually share different characteristics from other failed test cases (i.e., FPs).
For example, when inspecting the evaluation results in Section~\ref{sec:approach:rq2}, we find that bug-finding test cases are more likely to throw exceptions that are rarely or never thrown by FPs, and they are more likely to call a method that is rarely or never called by FPs.
One possible reason for this is that bugs are usually related to exceptional or corner cases instead of normal cases, and are rare compared to correctly implemented features.
Based on these observations, FPs and TPs can be regarded as normal samples and outliers, respectively, and the ranking problem can be converted into an outlier detection problem.
With this intuition, we build our ranking method based on outlier detection algorithms.

\begin{figure}[!t]
    \centering
    \includegraphics[width=0.47\textwidth]{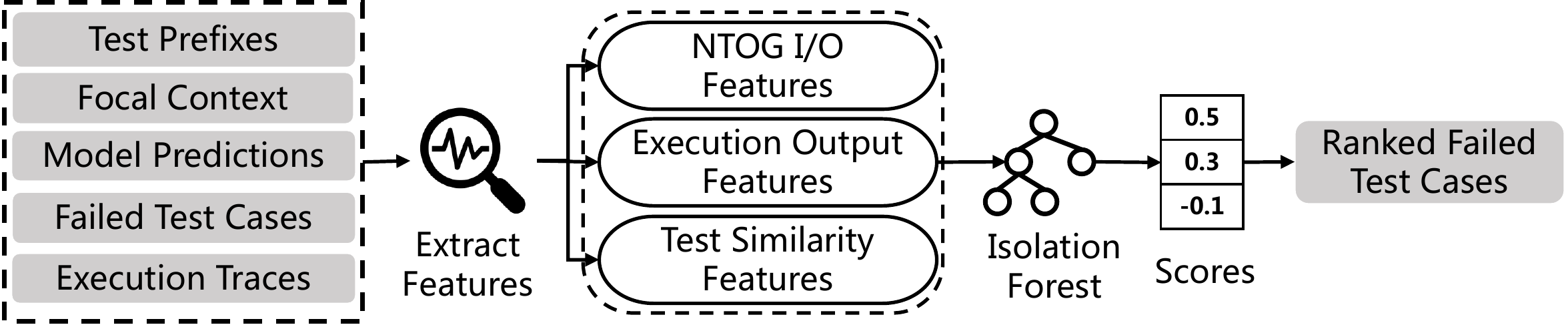}
    \caption{The procedure of our ranking method.}
    \label{fig:ranking}
    \vspace{-0.3cm}
\end{figure}

Figure~\ref{fig:ranking} presents the procedure of our ranking method.
Specifically, given a bug and the failed test cases generated by an NTOG approach for this bug, our ranking method first collects inputs from multiple sources, including the inputs of the NTOG approach, the predictions of the approach's neural model, the source code and the stack traces of the failed test cases.
Next, it extracts three dimensions of features for each failed test case.
Then, based on such features, the Isolation Forest~\cite{liu2008isolation} algorithm is adopted to detect outliers from all the failed test cases and calculate the anomaly score of each failed test.
Finally, we rank the failed test cases based on their anomaly scores.
The more likely a test case is to be an outlier, the higher its rank is.
Based on these ranked test cases, Found@K can be easily calculated.

\subsection{Features of Generated Test Cases}

\begin{table*}[!t]
\centering
\ra{1.1}
\caption{The features crafted for our ranking method}
\label{tab:features}
\begin{tabular}{|p{0.125\textwidth}|p{0.195\textwidth}|p{0.61\textwidth}|}
\hline 
\textbf{Feature Group} & \textbf{Features} & \textbf{Description} \\
\hline 
\multirow{4}{*}{NTOG I/O} & focal\_method\_name\_count & The number of the test cases of which the focal method name is the same as that of this testj.\\
\cline{2-3} 
 & test\_distinct\_code\_line & The number of the code lines in this test case that do not appear in other tests. \\
\cline{2-3} 
 & is\_exception & The generated oracle expects an exception. \\
\cline{2-3} 
 & is\_no\_exception & The generated oracle expects no exception. \\
\hline 
\multirow{6}{*}{Execution Output} & test\_prefix\_exception & Whether the regression behavior of the test prefix is throwing an exception. \\
\cline{2-3} 
 & trace\_exception\_count & The number of the test cases that have the same trace exception as this test case. \\
\cline{2-3} 
 & trace\_exception\_msg\_count & The number of the test cases of which the stack traces contain the same exception message as this test case. \\
\cline{2-3} 
 & is\_exp\_trace\_exception & Whether the trace exception is expected. The trace exception is regarded as expected if it is an AssertionFailError or its name appears in the focal method or the focal docstring. \\
\cline{2-3} 
 & unexp\_trace\_e\_count & This feature is set to 0 if this trace exception is expected, and to the number of the test cases of which the trace exceptions are unexpected and the same as this exception if this trace exception is unexpected. \\
\cline{2-3} 
 & focal\_unexp\_trace\_e\_count & This feature is set to 0 if the trace exception is expected. If this trace exception is unexpected, this feature is set to the number of the test cases of which the trace exceptions are unexpected and the focal methods as well as the trace exceptions are the same as those of this test case. \\
\hline 
Text Similarity & test\_doc\_sim & The textual similarity between the test case and the focal docstring. \\
\hline
\end{tabular}
\end{table*} 
Based on the characteristics of this ranking task, we manually craft three dimensions of features to help outlier detectors distinguish TPs and FPs, as shown in Table~\ref{tab:features}:

\textbf{NTOG I/O}: This dimension of features is extracted from the input and output of the NTOG approach.
For each failed test case, \textbf{focal\_method\_name\_count} and \textbf{test\_distinct\_code\_line} measure its uniqueness and rareness.
If this test case checks a focal method that is not checked by others or contains distinct code lines, it is rarer and more unique, and therefore is more likely to be a TP.
\textbf{is\_exception} and \textbf{is\_no\_exception} are used to indicate the type of the generated oracle.

\begin{figure}[!t]
    \centering
    \vspace{3mm}
\begin{lstlisting}[language=Java,linewidth={\linewidth},frame=tb,basicstyle=\small\ttfamily]
org.jfree.data.KeyedObjects2D_ESTest::test131
junit.framework.AssertionFailedError: expected:(@\textless1\textgreater@) but was:(@\textless2\textgreater@) 
   at org.junit.Assert.fail(Assert.java:88)
   org.junit.Assert.failNotEquals(Assert.java:743) 
   org.junit.Assert.assertEquals(Assert.java:118)
   ......
\end{lstlisting}
    \caption{The stack trace of a failed test case.}
    \label{fig:exec_trace}
    \vspace{-0.3cm}
\end{figure}

\textbf{Execution Output}: Before ranking the failed test cases, we have already obtained their stack traces.
Figure~\ref{fig:exec_trace} presents the stack trace of a failed test case.
We can see that a stack trace contains the qualified name of the test case (the first line), the exception line (the second line) and the execution stack (other lines).
The exception line consists of the raised exception, e.g., ``{\tt junit.framework.Assertio nFailedError}'', and its message, e.g., ``expected:\textless1\textgreater~but was:\textless2\textgreater''.
We refer to such raised exception as the trace exception and denote the last item of the trace exception, e.g., ``AssertionFailedError'' in Figure~\ref{fig:exec_trace}, as the trace exception name.
\textbf{test\_prefix\_exception} can help identify the source of the trace exception.
For each failed test case, we extract this feature by parsing its test prefix and checking whether there is any catch clause. 
\textbf{trace\_exception\_count} and \textbf{trace\_exception\_msg\_count} can reflect the uniqueness of this trace exception.
If the trace exception is an AssertionFailError, or its name appears in the focal method or the focal docstring, we regard the trace exception as an expected trace exception.
Because it is common and usually expected that a failed test is caused by assertion fails and that a generated test case throws an exception recording in the focal method or the focal docstring.
Intuitively, the more surprising the trace exception is, the more likely the test case reaches some unspecified or buggy behaviors.
Therefore, we use \textbf{is\_exp\_trace\_exception} to indicate whether the trace exception is expected and leverage \textbf{unexp\_trace\_e\_count} to measure the uniqueness of an unexpected trace exception among all the unexpected trace exceptions of this bug.
\textbf{focal\_unexp\_trace\_e\_count} further refines the measurement of such uniqueness by additionally considering the focal method for counting.

\textbf{Text Similarity}: \textbf{test\_doc\_sim} measures the textual similarity between the test case and the focal docstring.
Since the focal docstring may contain the informal specification of the focal method, a failed test case that is more similar to the focal docstring may be more likely to reveal bugs.
To calculate such similarity, we convert the focal method and the test case into TF-IDF vectors and calculate their cosine similarity.

In summary, because these features are crafted for outlier detection techniques, they either provide the basic information of each test case, e.g., is\_exception and is\_no\_exception, or try to measure the unexpectedness and rareness of each test case.

We would like to claim our ranking method should not be regarded as preferring exception oracles.
Although seven out of the eleven features seem related to exceptions, five out of the seven features focus on trace exceptions, which are obtained from the execution results instead of the source code of test cases.
Our ranking method aims to rank failed test cases.
Each failed test case, no matter whether its oracle is an exception or assertion oracle, has a stack trace and a stack exception.
Therefore, our method has no inherent preference for exception oracles.
In addition, all the features can be extracted from the inputs and outputs of an NTOG approach and are independent of NTOG approaches.

\subsection{The Adoption of Isolation Forest}
Our ranking method uses Isolation Forest to calculate anomaly scores.
Isolation Forest is a widely-used unsupervised technique for outlier detection.
It has a linear time complexity that has exhibited high accuracy over a variety of datasets~\cite{liu2012isolationbased}.
We implement Isolation Forest using the scikit\text{-}learn~\cite{pedregosa2011scikitlearn} toolkit with default parameters.
Since Isolation Forest is based on random partitioning of features, for each bug, we run iForest 10 times with different random states and report the mean value of each metric as the result of one experiment.

\subsection{Evaluation of Our Ranking Method}
To assess the effectiveness of our proposed ranking method and measure TOGA's bug-finding performance in a cost-effective way, we evaluate TOGA and NoException using TEval@buggy plus our ranking method and compare the evaluation results with those obtained by using TEval@buggy plus random ranking.
Similar to our ranking method, the random ranking method is also run 10 times with different random states in one experiment to reduce bias.
Found@K metrics with k=1, 3, 5, and 10 are reported to show whether our proposed ranking method can improve the practicability of TOGA.
Wilcoxon signed-rank tests~\cite{wilcoxon1992individual} at the confidence level of 95\% and Cliff's deleta~\cite{cliff2014ordinal} are also conducted.

\begin{table}[!t]
    \centering
\caption{The evaluation results of our ranking method.} 
\label{tab:rq4_result}
    \begin{threeparttable}
    \begin{tabular}{@{}llcccc@{}}
    \toprule
    \textbf{Approach} & \textbf{Ranking} & \textbf{F@1} & \textbf{F@3} & \textbf{F@5} & \textbf{F@10}\\
    \midrule
    \textbf{NoException} & \textbf{Random} & 11.81 & 22.91 & 27.76 & 35.93\\
    \textbf{NoException} & \textbf{Ours}  & 13.74 & 26.38 & 30.07 & 37.52 \\
    \midrule
    \textbf{TOGA} & \textbf{Random} & 9.89 & 20.85 & 27.61 & 39.09\\
    \textbf{TOGA} & \textbf{Ours} & \textbf{15.19} & \textbf{29.59} & \textbf{36.31} & \textbf{45.30}\\
    \bottomrule
    \end{tabular}
    \begin{tablenotes}
        \footnotesize
        \item All experiments uses TEval@buggy. F@K refers to Found@K.
    \end{tablenotes}
    \end{threeparttable}
    \vspace{-0.3cm}
\end{table}

The evaluation results are shown in Table~\ref{tab:rq4_result}.
We can observe that for both NoException and TOGA, our ranking method can improve their bug-finding performance in terms of Found@K, especially when K is small.
In detail, when using TOGA, our ranking method can help developers find 5.3 (53.6\%), 8.74 (41.9\%) and 8.70 (31.5\%) more bugs if developers only check the recommended top-1, 3 and 5 failed test cases of each bug.
The statistical analysis shows that the performance improvements of NoException after using our ranking method are significant in terms of all Found@K with at least medium effect sizes.
Our ranking method also significantly improves the Found@K of TOGA with large effect sizes.
These results indicate that our ranking method can effectively improve the practicability of both NoException and TOGA.

According to Table~\ref{tab:rq4_result} and our statistical analysis, we surprisingly observe that NoException+Random significantly performs better than TOGA+Random in terms of Found@1, and Found@3 with at least medium effect size, which means the cost-effectiveness of NoException is better than that of TOGA if developers only randomly check one or three failed tests cases to find a bug.
But NoException+Random does not significantly outperform TOGA+Random in terms of Found@5 and performs significantly worse in terms of Found@10.
This is reasonable because as presented in Section~\ref{sec:approach:rq3}, TOGA can generate more bug-finding test cases for more bugs, and if the search space is enlarged, it would be more likely to find a bug-finding test case generated by TOGA.

It is interesting that when using our ranking method, TOGA achieves better Found@K than NoException for all Ks.
All the p-values are less than 0.05, indicating significant performance differences.
These results demonstrate that when using a proper ranking method, TOGA can indeed be more practical than NoException, and also highlight the importance of ranking methods for evaluating NTOG approaches.

\section{Towards More Realistic Evaluation}\label{sec:pipeline}

\begin{figure*}[!t]
    \centering
    \includegraphics[width=0.90\textwidth]{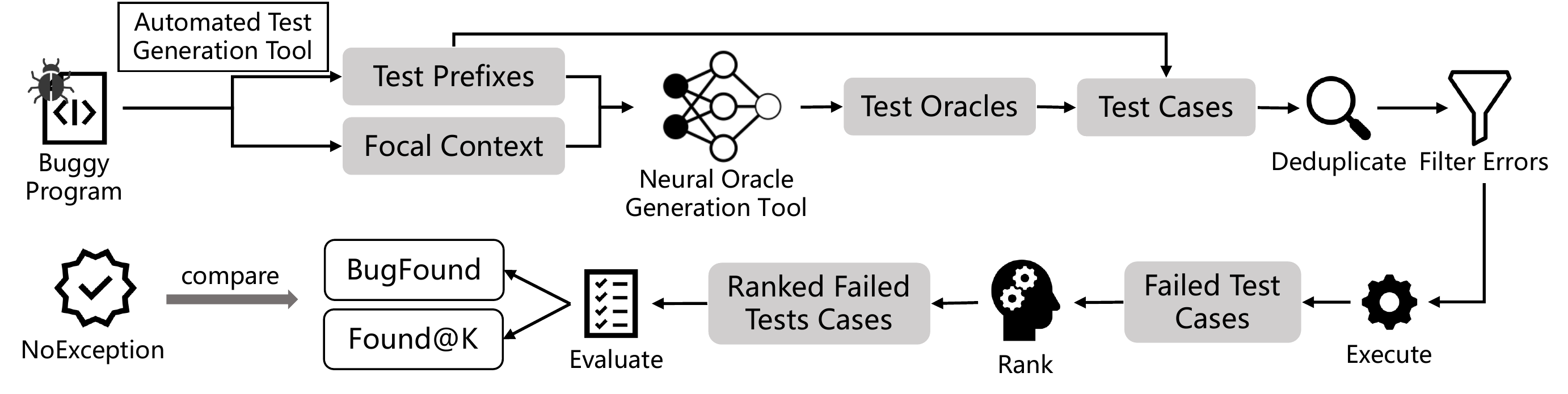}
    \caption{The procedure of TEval+.}
    \label{fig:approach}
    \vspace{-0.1cm}
\end{figure*}

In this section, we incorporate all the enhancements we made on TEval with TEval, and propose a more realistic evaluation method named TEval+ for NTOG.
Figure~\ref{fig:approach} presents the overall procedure of TEval+, which is similar to the procedure of TEval described in Section~\ref{sec:eval_methods}.
However, TEval+ makes several enhancements over TEval, which we summarize as seven rules of thumb to boost NTOG approaches towards their practical usages:
\begin{itemize}[leftmargin=*]
    \item {\bf Be cautious to avoid data leakage.} Some information contained in a bug benchmark is not available when applying NTOG approaches in practice, e.g., the patch fixing the target bug. Therefore, we should be careful not to use such information when constructing inputs for NTOG approaches. Following this rule, TEval+ uses EvoSuite to generate test prefixes on the buggy program versions instead of the bug-fixed program versions.
    \item {\bf Reduce random noise.} During the evaluation, the techniques relying on randomized algorithms should be run multiple times to reduce noise. In fact, this rule has been proposed by prior work~\cite{arcuri2011practical}, but not all NTOG work respects it. Following this rule, TEval+ runs EvoSuite 10 times and accordingly conducts experiments 10 times to reduce the bias. In addition, for our proposed ranking method, we also run Isolation Forest 10 times because it is also a randomized technique. 
    \item {\bf Be careful about duplicate data.} Duplicate test cases may introduce bias. It is necessary to avoid feeding duplicate inputs to NTOG approaches and to deduplicate test cases before execution. Following this rule, TEval+ explicitly deduplicates the generated test cases before executing them.
    \item {\bf Handle compilation errors properly.} When encountering compilation errors, an evaluation method should carefully filter the corresponding test cases or try to fix such errors, and should avoid affecting other test cases with no errors. Based on this rule, TEval+ fixes the over-filtering problem in the original implementation of TEval.
    \item {\bf Keep real-world usage scenarios in mind.} The goal of evaluation methods is to measure the real-world performance of NTOG approaches. Therefore, evaluation methods for NTOG should always keep real-world usage scenarios in mind. Based on this rule, TEval+ does not generate test prefixes from the bug-fixed program versions, introduces an additional ranking step to mimic the usage scenario, and proposes the Found@K metric to measure the bug-finding performance of NTOG approaches practically.
    \item {\bf Compare NTOG approaches with simple and straightforward baselines.} Potential users of NTOG approaches also care about the benefits and Return on Investment (ROI) of neural models, which can be demonstrated by comparing NTOG approaches with simple and straightforward baselines. Following this rule, TEval+ introduces a naive baseline named NoException and explicitly compares NTOG approaches with this baseline.
    \item {\bf Pay attention to post-processing.} Besides generating oracles, how to make use of the generated oracles in practice is also an important question. To demonstrate this, TEval+ proposes a novel unsupervised ranking method to rank the failed test cases and significantly improves the practicability of the underlying NTOG approaches.
\end{itemize}

\section{Discussion}\label{sec:discussion}
\subsection{TOGA's performance on different oracle types}

\begin{table}[!t]
    \setlength\tabcolsep{3.6pt}
    \centering
    \ra{1.1}
    \caption{TOGA's performance on different oracle types using TEval@buggy.} 
\label{tab:rq3_result_breakdown}
    \begin{threeparttable}
    \begin{tabular}{@{}lcccc@{}}
    \toprule
    \textbf{Oracle Type} & \textbf{BugFound} & \textbf{Precision} & \textbf{\#TP} & \textbf{\#FP} \\
    \midrule
    \textbf{expect\_no\_exception} &  39.7 & 0.92\% & 64.3 & 6889.9 \\
    \textbf{expect\_exception} & 10.3 & 0.39\% & 20.8 & 5323.9 \\
    \textbf{assertion}	& 18.2 & 0.16\% & 25.6 & 16885.6 \\
    \midrule
    \textbf{all} & 64.4 & 0.38\% & 110.7 & 29099.4 \\
    \bottomrule
    \end{tabular}
    \end{threeparttable}
    \vspace{-0.2cm}
\end{table}

 \begin{table}[!t]
    \setlength\tabcolsep{3.7pt}
    \centering
\caption{TOGA's performance on different oracle types using TEval+.} 
\label{tab:rq4_result_breakdown}
    \begin{threeparttable}
    \begin{tabular}{@{}llcccc@{}}
    \toprule
    \textbf{Approach} & \textbf{Ranking} & \textbf{F@1} & \textbf{F@3} & \textbf{F@5} & \textbf{F@10}\\
    \midrule
    \textbf{expect\_no\_exception} & Ours & 15.62 & 25.47 & 29.69 & 34.94\\
    \textbf{expect\_exception} & Ours  & 4.59 & 5.81 & 6.38 & 8.16 \\
    \textbf{assertion} & Ours & 3.48 & 7.94 & 9.98 & 12.30\\
    \midrule
    \textbf{all} & Ours & 15.19 & 29.59 & 36.31 & 45.30 \\
    \bottomrule
    \end{tabular}
\end{threeparttable}
    \vspace{-0.2cm}
\end{table}

The oracles generated by TOGA can be divided into three types: expect\_no\_exception, expect\_exception and assertion.
Following Dinella et al.~\cite{dinella2022toga}, we break down TOGA's performance on the three types of oracles for a deeper investigation.
For TEval@buggy, we split the constructed test cases into three groups based on their oracle types and calculate the BugFound and Precision for each group, as shown in Table~\ref{tab:rq3_result_breakdown}.
Please note that some bugs can be triggered by different types of oracles.
Thus, the sum of the first three BugFound values in Table~\ref{tab:rq3_result_breakdown} is greater than the number of the bugs found by TOGA (i.e., 64.4).
For TEval+, we also split the ranked failed test cases into three groups and calculate the Found@K metrics for each group, as shown in Table~\ref{tab:rq4_result_breakdown}.
From Table~\ref{tab:rq3_result_breakdown} and Table~\ref{tab:rq4_result_breakdown}, we can see that TOGA's performance is quite different on the three types of oracles.
Best performance is achieved on expect\_no\_exception oracles, which strengthens the motivation of using NoException as a baseline.
There are substantial margins between the performance on expect\_no\_exception oracles and the performance on the other two types of oracles, calling for more efforts on improving the generation of expect\_exception and assertion oracles.

A variant of TOGA, which only outputs the expect\_no\_exception oracles generated by TOGA, seems similar to NoException.
However, the TOGA paper does not provide explicit and comprehensive comparison between TOGA and this variant.
NoException also provides additional benefits:
First, it is independent of other NTOG approaches, does not use neural models, and thus is simpler and easier to use.
Second, NoException can find more bugs than this variant (44.4 v.s. 39.7).
Therefore, we suggest practitioners explicitly consider NoException as a baseline to better evaluate NTOG approaches.

\subsection{Threats to Validity}
The first threat to the validity of this study is that EvoSuite and Isolation Forest are randomized algorithms and are affected by chance.
To cope with this threat, we run EvoSuite 10 times with different random states for each bug, run Isolation Forest 10 times with different random states in each experiment, and report the mean value of each evaluation metric as the result.
In addition, we also conduct statistical analysis to demonstrate the significance of our observed performance differences.

Second, there may be errors and biases in our experiments.
To mitigate this threat, we re\text{-}use the trained model of TOGA provided by its authors, build TEval+ based on the official implementation of TEval, and follow the settings used by TOGA unless explicitly stated.
In addition, We have double-checked our code and data, and released them in our replication package~\cite{website:package}.

Another threat to validity is the potentially hidden gaps that are not realized by us between TEval+ and real-world usage scenarios.
A potential gap is that for each bug, both TEval and TEval+ use EvoSuite to generate test prefixes only for the classes that will be patched to fix this bug.
However, bug locations are unknown when developers use NTOG approaches to find bugs.
Therefore, this setting is potentially unrealistic.
We do not investigate this setting's impact in this work, because:
First, considering there are at least tens and hundreds of focal classes in a real-world project, generating test prefixes and oracles for all of them is expensive and time-consuming.
Second, there do exist some situations in practice where we only need to generate oracles for a limited number of classes or even methods.
For example, when using NTOG tools after a code commit, developers only need to generate test prefixes for the classes or even the methods that have been changed in this commit.
It would be interesting for future work to further investigate the impact of this setting on the bug-finding performance of NTOG approaches and to find and fill other gaps between the evaluation methods for NTOG and the real-world usage scenarios.

\section{Related Work}~\label{sec:related_work}
This section discusses the related work concerning automated test generation and test oracle generation.

\subsection{Unit Test Generation}
TEval leverages unit test generation tools to generate test prefixes for NTOG approaches.
Researchers have proposed many tools for automated unit test generation~\cite{pacheco2007randoop,fraser2012whole,sen2005cute,malburg2011combining,lukasczyk2022pynguin,zhu2022fuzzing,tufano2021unit}.
For example, EvoSuite~\cite{fraser2012whole,fraser2011evosuite} is a typical and widely used search-based test generation tool, which leverages genetic algorithms to optimize the coverage and the length of a whole test suite.
\textsc{Randoop}~\cite{pacheco2007randoop} is a random test generation tool, which generates unit tests by randomly selecting a method call to apply and finding the arguments for the call from previously\text{-}constructed objects.
Zhang et al.~\cite{zhang2011combined} proposed a tool name Palus, which enhances a random test generator with the information collected using static and dynamic analysis.
Symstra~\cite{xie2005symstra} is a constraint-based tool, which uses symbolic execution to explore the object states of an object-oriented system and generate tests.
DART~\cite{godefroid2005dart}, CUTE~\cite{sen2005cute} and Pex~\cite{tillmann2008pex} combine symbolic and concrete execution, i.e., use concolic execution, to enumerate feasible execution paths and synthesize test inputs.
Evacon~\cite{inkumsah2008improving} serially combines symbolic execution and search-based techniques to generate test cases with higher branch coverage.
Malburg and Fraser~\cite{malburg2011combining} proposed to use a constraint solver to help search-based techniques avoid being stuck in local optima.
Lukasczyk et al.~\cite{lukasczyk2022pynguin} proposed a test generation tool named \textsc{Pynguin}, which adapts the techniques behind EvoSuite and \textsc{Randoop} for Python projects.
In addition, fuzzers~\cite{miller1990empirical,zhu2022fuzzing} generates test inputs based on grammars or valid corpus, i.e., generation-based fuzzing, or by mutating seed inputs, i.e., mutation-based fuzzing.
Recently, Tufano et al.~\cite{tufano2021unit} proposed a learning-based generation tool named \textsc{AthenaTest} which pre-trains a BART model~\cite{lewis2020bart} using English and Code corpora, and fine-tunes this model to generate test cases with the corresponding focal context.

Different from the above-mentioned work, this work focuses on test oracle generation instead of unit test generation, and targets studying and improving existing evaluation methods for neural test oracle generation.

\subsection{Test Oracle Generation}

Although unit test generation tools are powerful, they still face the oracle problem~\cite{barr2014oracle} and cannot generate test cases that reveal functional bugs in the current program version.
To fill the gap, researchers have proposed some approaches to automatically generate test oracles that can capture functional bugs.
One type of them assumes that the intended behavior of a Unit Under Test (UUT) is informally described in its documentation, and synthesizes test oracles from the documentation based on natural language processing (NLP) techniques, manually-summarized rules or patterns, and search-based techniques~\cite{pandita2012inferring,tan2012tcomment,goffi2016automatic, blasi2018translating,blasi2021memo,zhai2020c2s}.
For example, 
Goffi et al.~\cite{goffi2016automatic} proposed Toradocu to generate exceptional oracles.
Toradocu first leverages NLP techniques to extract exceptions and their subjects and predicates from Javadoc, then creates conditional expressions based on pattern and lexical matching, and finally generates test oracles using run-time instrumentation.
JDoctor~\cite{blasi2018translating} enhances Toradocu by using not only pattern and lexical matching, but also semantic matching, and extends the crafted rules to generate preconditions, assertion oracles and exceptional oracles.
Blasi et al.~\cite{blasi2021memo} proposed Memo, which automatically derives metamorphic equivalence relations from natural language documentation.
The above-mentioned tools rely on specific patterns or rules and can hardly handle the flexibility and diversity of real-world code documentation.
C2S~\cite{zhai2020c2s} aligns words in comments and tokens in specifications and leverages a search-based technique to synthesize JML specifications~\cite{burdy2005overview} based on such alignment.
C2S does not rely on patterns but requires developer-written test prefixes to filter invalid specifications.
Moreover, prior work~\cite{dinella2022toga} showed that these tools struggle to infer bug-finding oracles for the Defects4J benchmark.

Recently, researchers proposed to leverage deep neural networks to generate test oracles by learning from developer-written test cases~\cite{watson2020learning,mastropaolo2021studying,dinella2022toga,mastropaolo2022using,tufano2022generating}, i.e., neural test oracle generation (NTOG).
Watson et al.~\cite{watson2020learning} proposed \textsc{Atlas}, which leverages a neural encoder-decoder model with copy network to generate assert statements based on developer-written test prefixes and focal methods.
Mastropaolo et al.~\cite{mastropaolo2021studying, mastropaolo2022using} pre-trained a T5 model~\cite{raffel2020exploring} using massive raw source code, abstracted source code and code comments, and fine-tuned this model for multiple code-related tasks including assert statement generation.
Tufano et al.~\cite{tufano2022generating} proposed to pre-trained a BART model~\cite{lewis2020bart} on large English and code corpora, and fine-tuned the pre-trained model to generate assert statements.
Recently, Dinella et al.~\cite{dinella2022toga} proposed an NTOG tool named TOGA, which is shown to achieve state-of-the-art bug-finding performance.

In this work, we conduct our case study using TOGA.
Different from existing studies, this work focuses on how to better evaluate the bug-finding performance of NTOG tools, instead of proposing a new test oracle generation approach.
Therefore, it is complementary to existing test oracle generation approaches and can inspire more realistic evaluation methods for this research field.

 \section{Conclusion and Future Work}\label{sec:conclusion}
This work focuses on the realistic evaluation for neural test oracle generation (NTOG).
We first point out three inappropriate settings in existing evaluation methods and comprehensively investigate their impacts on evaluating and understanding the bug-finding performance of NTOG approaches.
Our experimental results show that all three settings have negative impacts on measuring or understanding the performance of NTOG approaches, and even the state-of-the-art NTOG approach TOGA suffers from limited precision.
Next, we introduce a ranking step in existing evaluation methods and propose a more realistic evaluation metric Found@K to help measure the practicability of NTOG approaches.
Then, we instantiate this ranking step with a novel unsupervised ranking method, which is based on three dimensions of features specifically designed for this task and an outlier detector.
Our ranking method can significantly improve the cost-effectiveness of TOGA.
Finally, we incorporate all our enhancements with existing evaluation methods to construct a more realistic evaluation method named TEval+ and summarize seven rules of thumb to help practitioners better evaluate and understand the performance of NTOG approaches.

In the future, we plan to extend this work to consider more test prefix generation techniques, such as \textsc{Randoop} and EvoSuite with different fitness functions, and more neural test oracle generation approaches.

\begin{acks}
This research/project is supported by the National Natural Science Foundation of China (No. 62202420, No. 62172214 and No. U20A20173) and the Natural Science Foundation of Jiangsu Province, China (No. BK20210279). Zhongxin Liu gratefully acknowledges the support of Zhejiang University Education Foundation Qizhen Scholar Foundation.
\end{acks}

\balance
\bibliographystyle{ACM-Reference-Format}
\bibliography{reference}

\end{document}